\documentclass[twoside,twocolumn,9pt]{article}
\pdfoutput=1 

\usepackage{extsizes}
\usepackage[super,sort&compress,comma]{natbib} 
\usepackage[version=3]{mhchem}
\usepackage[left=1.5cm, right=1.5cm, top=1.785cm, bottom=2.0cm]{geometry}
\usepackage{balance}
\usepackage{widetext}
\usepackage{times,mathptmx}
\usepackage{sectsty}
\usepackage{graphicx} 
\usepackage{lastpage}
\usepackage[format=plain,justification=justified,singlelinecheck=false,font={stretch=1.125,small,sf},labelfont=bf,labelsep=space]{caption} 
\usepackage{float}
\usepackage{fancyhdr}
\usepackage{fnpos}
\usepackage[english]{babel}
\usepackage{array}
\usepackage{droidsans}
\usepackage{charter}
\usepackage[utf8]{inputenc}
\usepackage[usenames,dvipsnames]{xcolor}
\usepackage{setspace}
\usepackage[compact]{titlesec}
\usepackage{fixltx2e}
\usepackage{siunitx}
\usepackage{amssymb}

\usepackage{systeme}

\usepackage{cleveref}
\crefname{equation}{eq.}{eqs.}
\Crefname{equation}{Eq.}{Eqs.}

\crefname{figure}{fig.}{figs.}
\Crefname{figure}{Fig.}{Figs.}

\usepackage{epstopdf}

\definecolor{cream}{RGB}{222,217,201}
\graphicspath{ {./pics/} }
\DeclareGraphicsExtensions{.jpg,.jpeg,.png,.pdf,.tiff}

\begin{document}

\pagestyle{fancy}
\thispagestyle{plain}
\fancypagestyle{plain}{

\fancyhead[C]{\includegraphics[width=18.5cm]{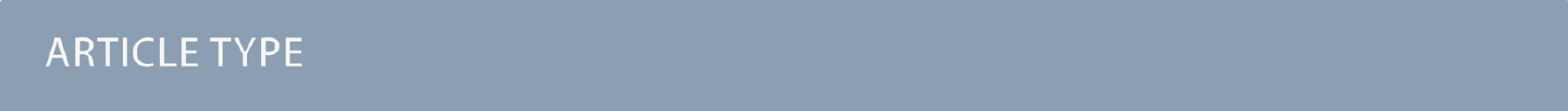}}
\fancyhead[L]{\hspace{0cm}\vspace{1.5cm}\includegraphics[height=30pt]{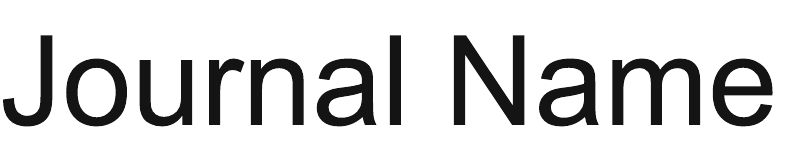}}
\fancyhead[R]{\hspace{0cm}\vspace{3.64cm}}
\renewcommand{\headrulewidth}{0pt}
}

\makeFNbottom
\makeatletter
\renewcommand\LARGE{\@setfontsize\LARGE{15pt}{17}}
\renewcommand\Large{\@setfontsize\Large{12pt}{14}}
\renewcommand\large{\@setfontsize\large{10pt}{12}}
\renewcommand\footnotesize{\@setfontsize\footnotesize{7pt}{10}}
\makeatother

\renewcommand{\thefootnote}{\fnsymbol{footnote}}
\renewcommand\footnoterule{\vspace*{1pt}%
\color{cream}\hrule width 3.5in height 0.4pt \color{black}\vspace*{5pt}} 
\setcounter{secnumdepth}{5}

\makeatletter 
\renewcommand\@biblabel[1]{#1}            
\renewcommand\@makefntext[1]%
{\noindent\makebox[0pt][r]{\@thefnmark\,}#1}
\makeatother 
\renewcommand{\figurename}{\small{Fig.}~}
\sectionfont{\sffamily\Large}
\subsectionfont{\normalsize}
\subsubsectionfont{\bf}
\setstretch{1.125} 
\setlength{\skip\footins}{0.8cm}
\setlength{\footnotesep}{0.25cm}
\setlength{\jot}{10pt}
\titlespacing*{\section}{0pt}{4pt}{4pt}
\titlespacing*{\subsection}{0pt}{15pt}{1pt}

\fancyfoot{}
\fancyfoot[CO]{\vspace{-7.1pt}\hspace{13.2cm}\includegraphics{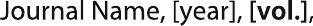}}
\fancyfoot[CE]{\vspace{-7.2pt}\hspace{-14.2cm}\includegraphics{RF}}
\fancyfoot[RO]{\footnotesize{\sffamily{1--\pageref{LastPage} ~\textbar  \hspace{2pt}\thepage}}}
\fancyfoot[LE]{\footnotesize{\sffamily{\thepage~\textbar\hspace{3.45cm} 1--\pageref{LastPage}}}}
\fancyhead{}
\renewcommand{\headrulewidth}{0pt} 
\renewcommand{\footrulewidth}{0pt}
\setlength{\arrayrulewidth}{1pt}
\setlength{\columnsep}{6.5mm}
\setlength\bibsep{1pt}

\makeatletter 
\newlength{\figrulesep} 
\setlength{\figrulesep}{0.5\textfloatsep} 

\newcommand{\topfigrule}{\vspace*{-1pt}%
\noindent{\color{cream}\rule[-\figrulesep]{\columnwidth}{1.5pt}} }

\newcommand{\botfigrule}{\vspace*{-2pt}%
\noindent{\color{cream}\rule[\figrulesep]{\columnwidth}{1.5pt}} }

\newcommand{\dblfigrule}{\vspace*{-1pt}%
\noindent{\color{cream}\rule[-\figrulesep]{\textwidth}{1.5pt}} }

\makeatother

\twocolumn[
  \begin{@twocolumnfalse}
\vspace{3cm}
\sffamily
\begin{tabular}{m{4.5cm} p{13.5cm} }


\includegraphics{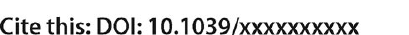} & \noindent\LARGE{\textbf{Drop-on-coilable-fibre systems exhibit negative stiffness events and transitions in coiling morphology$^\dag$}} \\
\vspace{0.3cm} & \vspace{0.3cm} \\

 & \noindent\large{Herv\'e Elettro$^{\ast}$\textit{$^{a,b}$}, Fritz Vollrath\textit{$^{c}$}, Arnaud Antkowiak\textit{$^{a,d}$}, and S\'ebastien Neukirch\textit{$^{a}$}}\\

\includegraphics{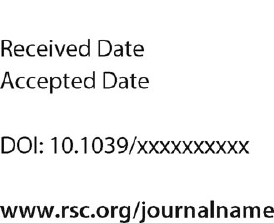} & \noindent\normalsize{We investigate the mechanics of elastic fibres carrying liquid droplets. In such systems, buckling may localize inside the drop cavity if the fibre is thin enough. This so-called drop-on-coilable-fibre system exhibits a surprising liquid-like response under compression, and a solid-like response under tension. Here we analyze this unconventional behavior in further details and find theoretical, numerical and experimental evidences of negative stiffness events. We find that the first and main negative stiffness regime owes its existence to the transfer of capillary-stored energy into mechanical curvature energy. The following negative stiffness events are associated with changes in the coiling morphology of the fibre. Eventually coiling becomes tightly locked into an ordered phase where liquid and solid deformations coexist.} \\

\end{tabular}

 \end{@twocolumnfalse} \vspace{0.6cm}

  ]

\renewcommand*\rmdefault{bch}\normalfont\upshape
\rmfamily
\section*{}
\vspace{-1cm}

\footnotetext{\textit{$^{\ast}$~corresponding author; e-mail: herve.elettro@usach.cl}}
\footnotetext{\textit{$^{a}$~Sorbonne Universit\'es, UPMC Univ Paris 06, CNRS, UMR 7190 Institut Jean Le Rond d'Alembert, F-75005 Paris, France;}}
\footnotetext{\textit{$^{b}$~Departamento de F\'{\i}sica, Universidad de Santiago de Chile, Av. Ecuador 3493, Santiago, Chile;}}
\footnotetext{\textit{$^{c}$~Oxford Silk Group, Zoology Department, University of Oxford, UK;}}

\footnotetext{\textit{$^{d}$~Surface du Verre et Interfaces, UMR 125 CNRS/Saint-Gobain, F-93303 Aubervilliers, France}}

\footnotetext{\dag~Electronic Supplementary Information (ESI) available: Video of numerical simulations of a drop-on-coilable-fibre system with $f_\gamma = 41$, showing the near and far-from-threshold mechanical response and related in-drop coiling rearrangement. See DOI: 10.1039/b000000x/}



%
%
%
%




%
%
%
%
%
%
\section{Introduction}
%
%
%
%
%
%
%
%
%
%


%
%

In systems where energy has been stored\cite{Thompson1979} or is continuously supplied\cite{Lakes2001}, force and deformation may operate in opposite directions, resulting in a so-called negative stiffness.
Active biological materials\cite{Martin2000} or systems under fluid loading\cite{Thompson1982} have been shown to exhibit such a behavior.
Locally negative stiffness appears as soon as the force-displacement curve of a system in non-monotonic. For example, it is encountered during the buckling of structures in specific geometries such as cylindrical shells\cite{Timoshenko1983,Hunt2011}, beams on elastic foundations\cite{Hunt1993}, elastic ribbons\cite{Fosdick2015}, or metamaterials\cite{coulais2015discontinuous}.

Here we show that drop-on-coilable-fibre systems also experience regimes of negative stiffness, mainly where capillary-stored energy is transferred into mechanical energy. Drop-on-fibre systems have a long history, from the textile industry\cite{Adam1937} to the coating of glass fibres\cite{Carroll1976,Quere1999}. Other examples include the wetting of fibres networks \cite{Lorenceau2006Wetting-of-Fibers} or the influence of capillary forces on fibre elastic deformation\cite{Sauret2017,Duprat2012}.
Elastocapillarity\cite{Roman2010}, the investigation of the deformation of elastic materials and structures by surface tension, lies at the interface between Physics\cite{Andreotti2016} and Engineering\cite{Syms2003} and is used as a way to functionalize and design new systems and materials\cite{Pineirua2010Capillary-origami,Leong2007Surface-Tension-Driven}.
%
%
\begin{figure}[ht]
\begin{center}
\includegraphics[width=0.85\columnwidth]{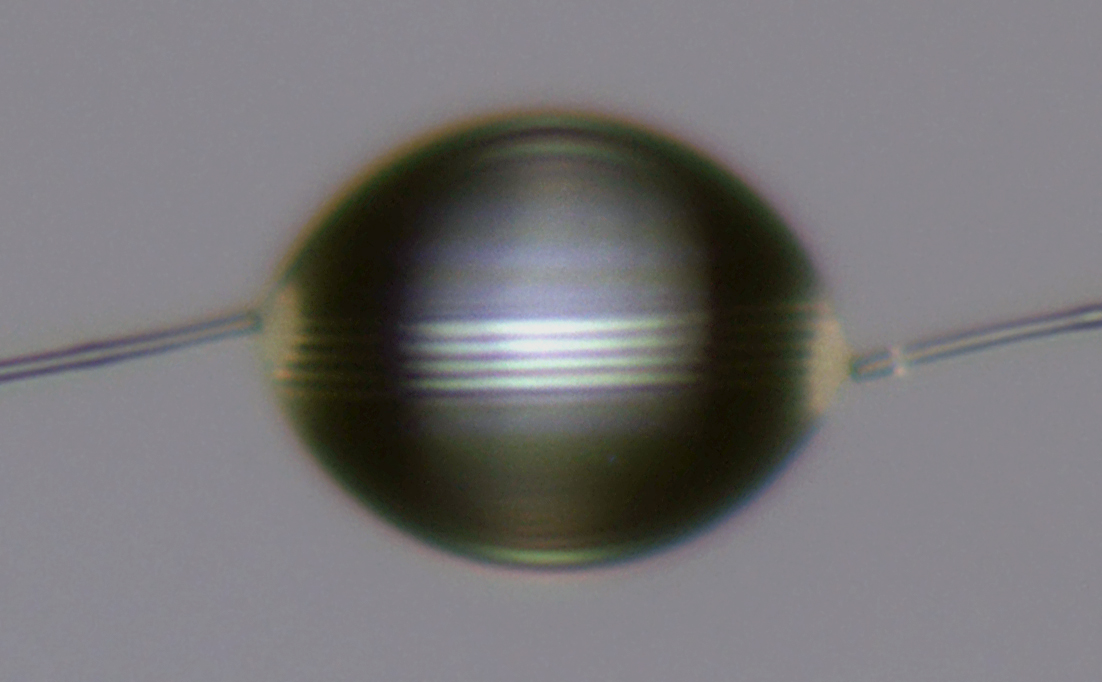}
\caption{Thin fibres may buckle and coil inside a liquid drop. A soft thermoplastic PolyUrethane fibre (Young modulus $E = 17$MPa) with radius $r=2.75\mu$m is spooled within a silicone oil drop of diameter $D=191\mu$m \cite{Elettro2015}.}
\label{coiling}
\end{center}
\end{figure}
%
%
%
%
Recently, taking inspiration from spider-silk fibres, we demonstrated that sufficiently thin fibres can locally buckle and coil within liquid drops, see \cref{coiling} and \citet{Elettro2015}. We further took advantage of the phenomena to design a highly extensible drop-on-coilable-fibre system.
In the following we show that in such a system, the active contribution of wetting energy gives rise to a subcritical buckling transition during which the stiffness of the system is negative.
We further investigate the consequences of subcriticality and show that it generates hysteresis in the mechanical response of the system (Section \ref{sec:subcritical}).
Additionally we show that the plateau tension contains coherent oscillations that consist of alternating regimes of positive and negative stiffnesses. These regimes are correlated with the drop deformation and the coiling morphology.
Packing morphology of a filament in a cavity\cite{Bayart2011} has been studied as a model for DNA viral capsids\cite{Purohit2003,Leforestier2009}. Morphogenesis of filaments in flexible cavities is also relevant in several biological systems\cite{Vetter2014}. \citet{Stoop2011} studied experimentally and numerically the packing of thin wires in spherical cavities, followed by \citet{Vetter2014}, who have recently shown that an ordered-to-disordered transition may occur for a filament in an elastic cavity by changing the confinement flexibility\cite{Vetter2014,Vetter2015}. 
\citet{Schulman2017} investigated the bending of microfibres around liquid droplets, while \citet{Roman2010} studied the deformation of the liquid drop as the rod is wrapping around it with one end free, thus restricting to ordered coiling through the release of twist\citep{Stoop2011}. 
Here we show that the packing morphology of the drop-on-coilable-fibre systems changes as the in-drop fibre length is increased and switches from disordered to ordered (Section \ref{sec:morphology}).

%
%
%
%
%
%
%
%
%

%
%
%
%
\section{A subcritical buckling transition} \label{sec:subcritical}
%
%
%
Subcritical transitions are discontinuous transitions, as opposed to supercritical transitions where the order parameter grows smoothly from zero. Subcritical transitions are present in a number of fields ({\em e.g.} Turing bifurcation in reaction-diffusion systems\cite{Brena-Medina2014}, transition to turbulence\cite{godreche2005}).  They imply dependance on the loading history, {\em i.e.}   hysteresis.
Buckling transitions may either be supercritical or subcritical depending on geometry: buckling of cylindrical shells\cite{Timoshenko1983} or ribbons\cite{Fosdick2015} are subcritical (discontinuous), while thin plates and slender beams buckle supercritically\citep{Gordon2003}. Metabeams may also buckle subcritically due to strong nonlinearities in their mechanical response\cite{coulais2015discontinuous}.
In this section we show that drop-on-coilable-fibre systems experience subcritical buckling as part of their unique conformation: in-drop buckling involves both the transfer of wetting energy into mechanical energy and the presence of a non-constant system length, which is reminiscent of beam buckling in sliding sleeves\cite{Bigoni2015}.
\begin{figure}[ht]
\begin{center}
\includegraphics[resolution=300,width=0.45\textwidth]{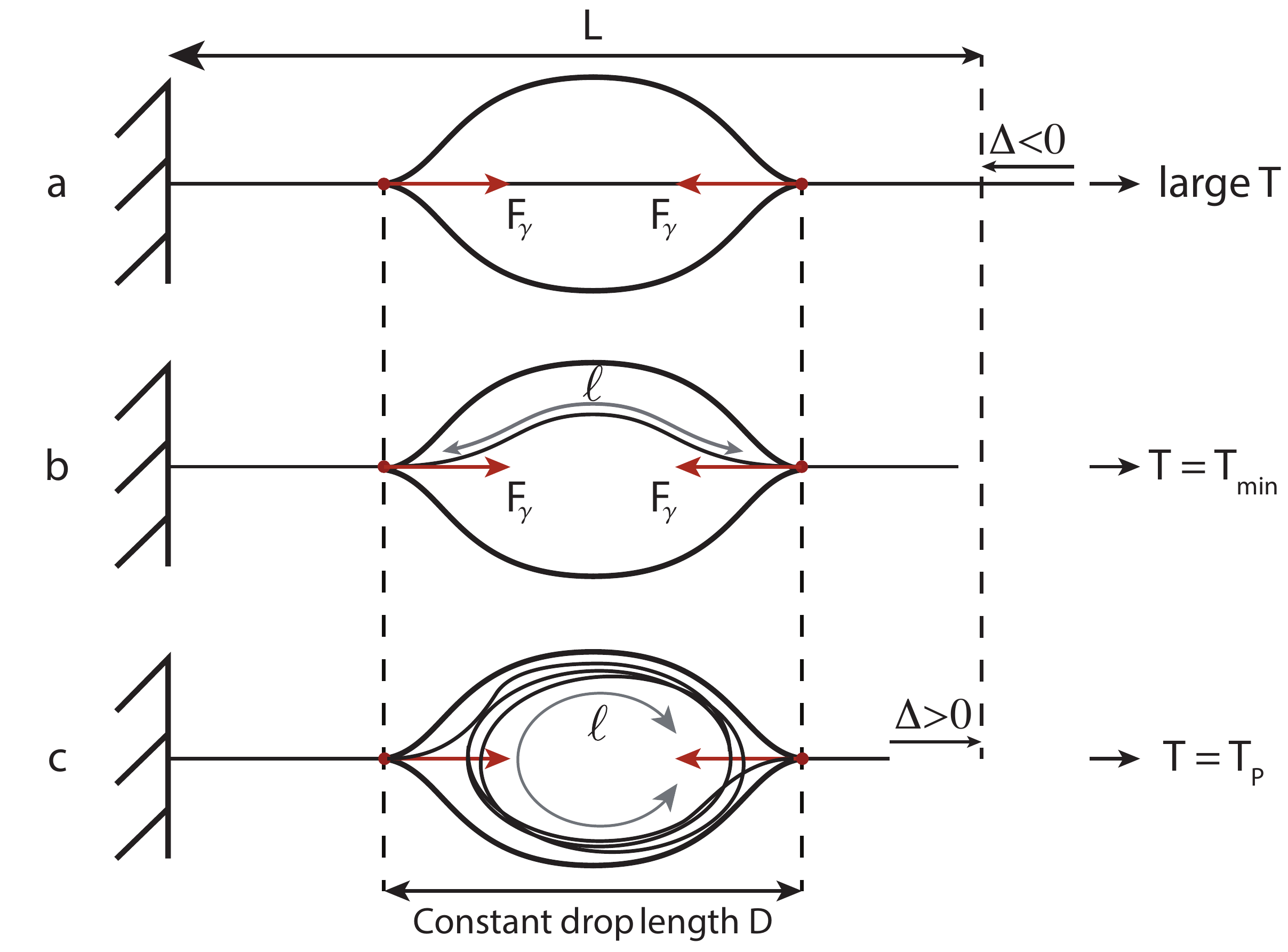}
\caption{\label{subcritical_sketch}The subcritical nature of the coiling mechanism derives from the specific mechanics of in-drop capillary buckling. (a): The fibre is under large tension, preventing it from coiling locally. (b): The tension reaches a minimum $T_\text{min}$ and buckling begins. The drop retains its size $D$, while swallowing the fibre. (c): The in-drop fibre length $\ell$ over which coiling occurs increases from $D$ (in case (a)) to $D+\Delta$ (in cases (b) and (c)), which decreases the system resistance.}
\end{center}
\end{figure}			

As shown in \citet{Elettro2015}, the drop-on-coilable-fibre system displays a classic solid-like behavior in tension and a remarkable liquid-like behavior in compression. 
In the solid-like regime, see \cref{subcritical_sketch}(a), the system behaves like a spring, that is the applied  tension $T$ is linearly related to the elongation $|\Delta|$ through the Young's modulus $E$ of the fibre, $T=EA |\Delta|/L$, where $A$ is the area of the cross section of the fibre, and $L$ the length of the fibre in its rest state. 
In the liquid-like regime, \cref{subcritical_sketch}(c), the system behaves like a soap film, that is 
it adapts its length while staying in a state of constant tension $T \approx T_P$ with\cite{Elettro2015}
\begin{equation}\label{equa:T-analytic}
T_\text{P}  = F_\gamma - \frac{1}{2} \pi E \frac{r^4}{D^2}
\end{equation}
where 
$F_\gamma = 2 \pi r \,\gamma \, \cos \theta_Y$ is the meniscus force,
$\gamma$ the liquid-vapor interface energy, $\theta_Y$ the contact angle of the liquid on the solid, $r$ the radius of circular cross-section of the fibre, $I=\pi r^4/4$ its second moment of area ($EI$ being the fibre resistance to bending), and $D$ is the drop length, measured as the meniscus-to-meniscus distance.
For large drops on small fibres, the drop length $D$ is close to the (almost spherical) drop diameter\citep{Carroll1976}. In the considered range of drop/fibre sizes, the departure from sphericity does not exceed $10\%$ and justifies a spherical drop assumption at first order.
Eq.~\ref{equa:T-analytic} results from an analysis in the moderate post-buckling regime and can be interpreted in terms of energies: per unit length of wetted fibre, the left-hand side is the work of the external applied tension while the right-hand side is the difference between the wetting and the bending energies (coiling being favorable when this difference is positive).
\subsection{Force undershoot during in-drop buckling}
%
%
%
We then consider a spherical drop of diameter $D$ resting on a long fibre of length $L \gg D$.
For an given capillary force $F_\gamma$, a typical in-drop buckling experiment starts with a drop resting on a fibre held taut under large tension $T$, \cref{subcritical_sketch}(a). As we gradually decrease $T$, the fibre remains straight until $T$ reaches $T_\text{min}$ and the portion of fibre inside the drop buckles, \cref{subcritical_sketch}(b).
Inside the drop the fibre is subject to a compressive force $P = F_\gamma - T$, and as tension $T$ is decreased, compression $P$ increases until buckling is reached. The buckling threshold $P_\text{buck} = F_\gamma - T_\text{min}$ depends on the capillary force $F_\gamma$ and detailed calculations yield\cite{Elettro2015_theory}
\begin{equation}\label{undershoot_eq}
\sqrt{\left(F_\gamma - T_{\text{min}} \right) \frac{D^2}{EI}}
+ \sqrt{ \frac{T_{\text{min}}  D^2}{EI} } \; \tan\left[ \frac{1}{2} \, \sqrt{\left(F_\gamma - T_{\text{min}} \right) \frac{D^2}{EI}} \, \right] = 0
\end{equation}
where we see that the portion of fibre that buckles has length $D$, that is buckling tends to be localized inside the drop.
We note that for low capillary force $F_\gamma$, we have $T_{\text{min}}=0$ as $F_\gamma = \pi^2 \, EI/D^2$, while for large capillary forces $F_\gamma \gg EI/D^2$ we have $T_\text{min} \simeq F_\gamma -4\pi^2 EI/D^2$.

\begin{figure}[ht]
\begin{center}
\includegraphics[resolution=300,width=0.45\textwidth]{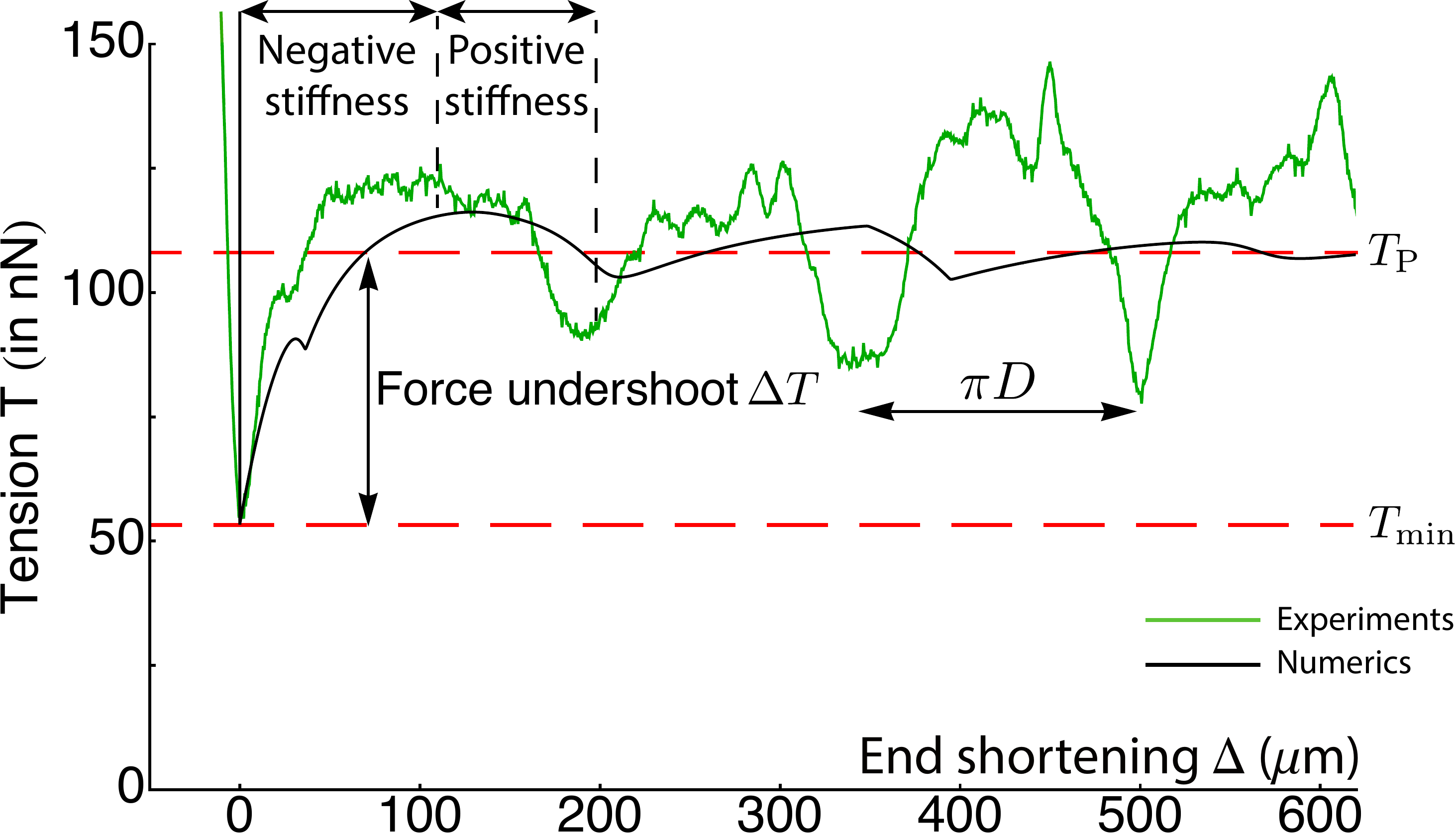}
\caption{Drop-on-coilable-fibre systems present an alternance of regimes of positive and negative stiffnesses. Comparison between numerical computations (black line) and experimental data (green line) on a TPU/silicone oil system (drop length $62\pm 2 \mu$m, TPU fibre diameter $1.9\pm 0.4\mu$m and dimensionless capillary force $f_\gamma = 41\pm20 > f_\gamma^{th} =\pi^2$, see Section \ref{ref:hysteresis}). The one coil period and the initial negative stiffness regime are well reproduced by the numerical model. Adapted from our previous work in \citet{Elettro2015}.}
\label{fig:Undershoot}
\end{center}
\end{figure}

As buckling grows, additional fibre length enters the drop by sliding along the meniscii in such a way that the drop length $D$ ({\em i.e} the distance between the two meniscii) remains constant. This increase in in-drop fibre length $\ell$ yields a decrease in bending resistance of the system. We compute the behaviour of the system in this incipient buckling regime.  
For simplicity reasons, we work under the assumption that the capillary force $F_\gamma$ is large compared to the bending force $EI/D^2$, in which case the drop meniscii can be viewed as sliding clamps\cite{Elettro2015_theory}.
When a clamped beam of length $\ell$ is buckling under a compressive force $P$, it undergoes an end-shortening $\Delta$ and in the incipient buckling regime (small $\Delta/\ell $) we have\cite{Bazant2010,Neukirch2014}
\begin{equation} \label{eq:supercrit}
P\ell^2/EI = 4\pi^2 + 2\pi^2 \, \frac{\Delta}{\ell}
\end{equation}
In the drop-on-fibre experiment, \cref{subcritical_sketch}, the compressive force is $P = F_\gamma - T$ and the length of the system under buckling is $\ell = D + \Delta$. Replacing $\ell$ and developping \cref{eq:supercrit} for small end shortening $\Delta/D \ll 1 $ yields
\begin{equation} \label{eq:subcrit}
\frac{P D^2}{EI}= (F_\gamma -T)\frac{D^2}{EI} \simeq 4\pi^2 - 6\pi^2 \, \frac{\Delta}{D}  
\end{equation}
\Cref{eq:supercrit,eq:subcrit} show that while classical buckling is supercritical (positive stiffness $\partial P/\partial \Delta$ in \cref{eq:supercrit}), in-drop buckling is subcritical (negative stiffness $\partial P/\partial \Delta$ in \cref{eq:subcrit}). This change in nature is due to the non-constant fibre length over which buckling occurs.
Consequently, as the fibre buckles in the drop, \cref{subcritical_sketch}(b), \cref{eq:subcrit} shows that $T$ is {\em increasing} with the end-shortening $\Delta$: the tension is reduced to $T=T_\text{min}$ to trigger buckling and is then expected to shoot back upwards after buckling. 
Eventually the tension stabilizes on the plateau value $T_\text{P}$ given by \cref{equa:T-analytic}, with $T_\text{P} > T_\text{min}$, see \cref{subcritical_sketch}(c) and \cref{fig:Undershoot}.
The presence of a drop thus modifies the nature of the buckling transition from supercritical (in classic slender beams/fibres) to subcritical (in drop-on-coilable-fibre systems).
This behavior is deeply related to the fact that the drop-on-coilable-fibre system is a liquid-solid hybrid: the meniscii act as fixed sliding sleeves and force additional fibre to enter the drop, exchanging wetting energy into curvature energy.

Several factors may influence the buckling threshold, for example the applied end rotation or the weight of the drop. End rotation is prevented by holding the fibre at one extremity while letting the other end free for a few seconds before attachment, in order to relax twist. The influence of twist, that builds up during fibre coiling, is analyzed in \cref{sec:morphology}. The influence of the drop  weight $M\, g$ is characterized by comparing it to the capillary force $F_\gamma$: introducing $C_\text{grav} = \tfrac{Mg}{2 F_\gamma}$, we have $C_\text{grav}< 2\%$ for drop/fibre sizes used in the present work. The effect of gravity is analyzed in more details in \citet{Elettro2015phd}.

In \cref{fig:Undershoot}, we show an experimental force-displacement curve of a drop-on-coilable-fibre system, along with a comparison with numerical simulations.
The numerical simulations model the fibre as an elastic filament, obeying Kirchhoff equilibrium equations\cite{Audoly2010}. The filament is in interaction with a spherical drop with two compressive point forces at the meniscii locations and a soft-wall barrier potential forbidding exit at any other point. The equilibrium of the system is solved using two-points boundary-value problem techniques (shooting method in Mathematica, and collocation method using the Fortran AUTO code). Note that the weight of the drop and the self-contact of the filament are not taken into account in the model. The numerical simulations only use externally measured parameters ({\em e.g.} drop length, fibre radius, surface tension) and no fitting parameter.
Sensor drifting and force offset during deposition of the drop imply that our experimental force data lacks an absolute reference. Consequently we globally adjust it so that the average value of the plateau tension corresponds to \cref{equa:T-analytic} (see Materials and Methods section).
Figure~\ref{fig:Undershoot} shows that the model recovers the force undershoot, the presence of negative and positive stiffnesses intervals, and the periodicity in end-shortening seen in the experiments.
%

\begin{figure}[ht]
\begin{center}
\includegraphics[width=0.85\columnwidth]{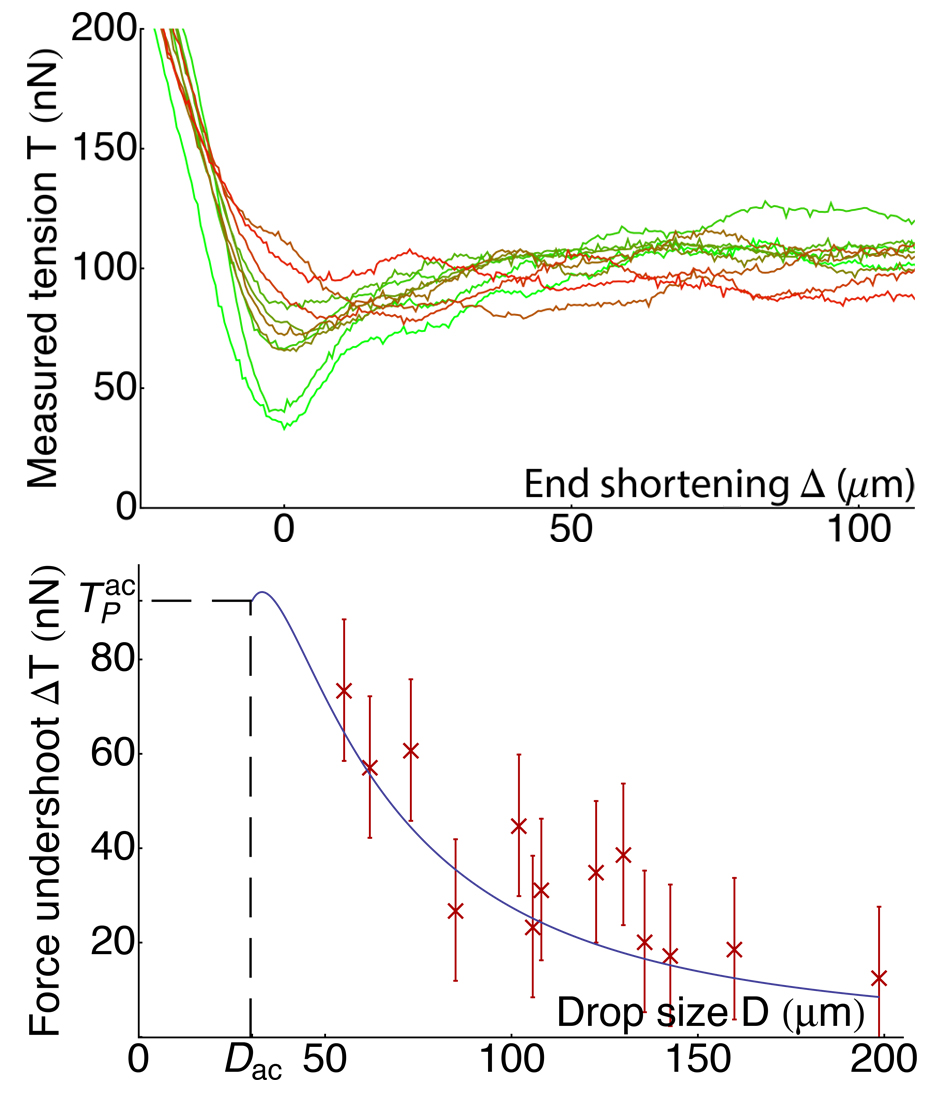}\\
\caption{Top: Force-displacement curves for different drop sizes on a $2 \pm 0.4 \mu$m diameter fibre. Colors correspond to dimensionless capillary forces $f_\gamma$, from $28$ (green) to $236$ (red). 
Bottom: Force undershoot as a function of drop length $D$ for the same fibre. The theoretical prediction, solid line, is in excellent agreement with the experimental data considering the absence of any fitting parameter. 
The value $T_\text{P}^{\text{ac}}$ is calculated from \cref{equa:T-analytic} with $F_\gamma$ such that \cref{undershoot_eq} is fulfilled with $T_\text{min}=0$: $F_\gamma=\pi^2 EI/D^2$.
}
\label{Undershoot_vs_Lw}
\end{center}
\end{figure}

On the one hand, the combination of \cref{equa:T-analytic,undershoot_eq} yields theoretical values for the force undershoot $\Delta T= T_\text P - T_\text{min}$, and on the other hand it is measured experimentally as the difference between the minimum force and the average value of the plateau tension.
This definition of the force undershoot as a relative quantity eludes the problem of sensor drifting exposed above. For each drop/fibre couple, the undershoots are measured three times to ensure reproducibility. We record force-displacement curves for different drop lengths $D$, and we extract the force undershoot from each curve, see \cref{Undershoot_vs_Lw}.
When the system is subjected to a global compressive force $T<0$, global buckling occurs if the compression $-T$ exceeds a threshold $\sim EI/L^2$. As $L$ is large, this threshold is indeed very low (piconewtons for centimetric soft microfibres) and we conclude that the drop-on-fibre system is virtually unable of sustaining any global compressive force ($T<0$).
Consequently in-drop buckling is not possible as soon as parameters are such that $T_\text{min}<0$. For a given fibre, $F_\gamma$ is known and positive and a critical drop length $D_\text{ac}$ can be computed by setting $T_\text{min}=0$ in \cref{undershoot_eq}: $D_\text{ac}=\pi \; \sqrt{EI/F_\gamma}$, with drops of length $D<D_\text{ac}$ being unable to initiate buckling of the fibre.
At $D=D_\text{ac}$, $T_\text{min}=0$ and the force undershoot is maximum and equal to the plateau value $T_\text{P}$ given by \cref{equa:T-analytic} with $F_\gamma=\pi^2 \, EI/{D^2_\text{ac}}$, that is $T_\text{P}^\text{ac}=(\pi^2-2) \, EI/{D^2_\text{ac}}$, or $T_\text{P}^\text{ac}=(1-2/\pi^2 ) \, F_\gamma$.
As $D$ is increased the force undershoot decreases as $1/D^2$.
Fig.~\ref{Undershoot_vs_Lw} shows the excellent agreement between the analytical and the experimental values, especially considering the absence of any fitting parameter. 

All force measurements shown so far are displacement-controlled experiments. In the following, we shall consider force-controlled experiments. Due to the specific force signature of drop-on-coilable-fibre systems, the two types of experiments lead to different behaviours.

\subsection{Mechanical hysteresis}
\label{ref:hysteresis}
\begin{figure}[ht]
\begin{center}
\includegraphics[resolution=300,width=0.45\textwidth]{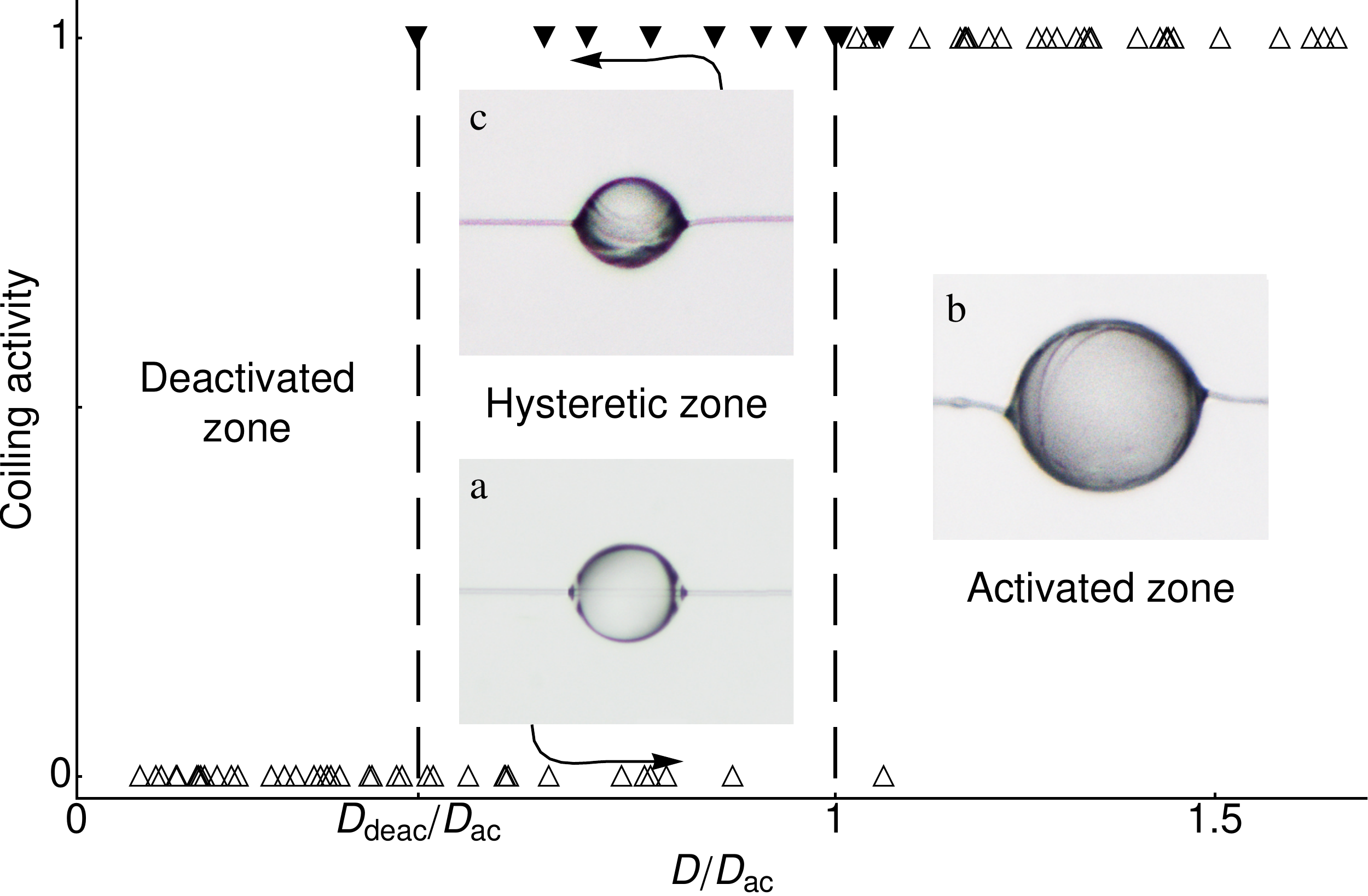}
\caption{ Coiling activity as a function of the dimensionless drop length $\tfrac{D}{D_{ac}}$ ($=\sqrt{\tfrac{f_\gamma}{\pi^2}}$). Transitions for activation and deactivation are different, displaying a mechanical hysteresis. Empty triangles represent experiments at increasing drop size while inverted filled triangles are for evaporating drops, decreasing in size. The fibre is made of TPU and measures $5.8 \pm 0.4 \mu$m in diameter. The scale is the same on all pictures. Inset (a) show a drop sitting on a straight fibre, while insets (b) and (c) show a coiled system during evaporation: in (b) the drop size $D=165 \mu$m in length, whereas in (c) $D=103 \mu$m.}
\label{Windlass_hysteresis}
\end{center}
\end{figure}

Here we perform experiments where the drop length $D$ is varied. This is equivalent to force-controlled experiments since the dimensionless force $f_\gamma = F_\gamma D^2/ EI \propto D^2$ is the control parameter for a given fibre.
%
%
We consider drop-on-fibre systems held in a very low state of tension $T \ll EI/D^2$, and report the coiling activity as a function of the drop length $D$. For small $D$, $f_\gamma$ is too small to induce coiling but, as drops with increasing $D$ are considered, in-drop buckling is observed as soon as $f_\gamma$ exceeds $\pi^2$ (see Eq.~\ref{undershoot_eq} with $T_\text{min}=0$ and $F_\gamma > 0$), that is we have {\em activation} of in-drop coiling if $D > D_\text{ac}$ with 
\begin{equation}
 D_\text{ac} = \sqrt{\frac{\pi^2 E r^3}{8\gamma\cos\theta_Y}}
\end{equation}
Experiments with increasing $D$ values are reported on \cref{Windlass_hysteresis}, where the coiling activity is plotted against the drop length $D$, and where we clearly see the activation threshold lying at $D/D_\text{ac}=1$. Coiling activity is defined as
zero when coiling is not possible and one when it is. Indeed, when the wetting energy overcomes the curvature energy, coiling becomes energetically favorable and will not stop until the drop is filled, case that has not been reached within our experimental range, even after coiling several cm of fibre in a 100 \si{\micro\meter} drop.
Once activated, the drop-on-coilable-fibre system is in a state of constant tension, the  plateau tension $T_\text{P}$ being given by Eq.~\ref{equa:T-analytic}. For such a coiled system, if we now decrease the drop length $D$ while keeping other parameters fixed, the plateau tension $T_\text{P}$ is going to decrease. As the system is virtually unable of sustaining any global compressive force ($T<0$), $T_\text{P}$ has to remain positive, and we anticipate a coiling {\em deactivation} when $T_\text{P}=0$, that is $D < D_\text{deac}$ with
\begin{equation}\label{deactivation_radius_eq}
D_{\text{deac}} = \frac{D_{\text{ac}}}{\sqrt{\pi^2/2}} = \sqrt{\frac{E r^3}{4\gamma\cos\theta_Y}}
\end{equation}
The deactivation drop-length is thus smaller than the activation length by a factor $\tfrac{\pi}{\sqrt{2}} \simeq 2.2$, resulting in two different thresholds.
We experimentally study the deactivation of in-drop buckling by use of evaporation. An ethanol microdrop-laden mist is sent onto a TPU fibre in a confinement chamber. Ethanol drops large enough to induce coiling are deposited on a TPU fibre. After coiling is achieved, the ethanol mist flow rate is slowly decreased, so that drops evaporate in a quasistatic manner. Quasistatic is defined in reference to the timescale for coiling rearrangement, which is here on the order of hundreds of milliseconds. The drop size is recorded optically throughout evaporation.

\begin{figure}[ht]
\begin{center}
\includegraphics[resolution=300,width=0.47\textwidth]{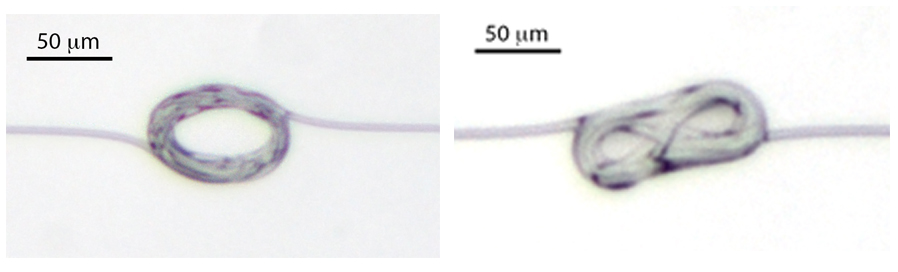}
\caption{The evaporation of a droplet containing a coiled fibre leaves highly regular dry coils (left). Residual end rotation may lead to twist instabilities, forming a lemniscate-shaped coil (right).}
\label{microcoil}
\end{center}
\end{figure}

We thus start with a coiled system in the activated zone $D>D_\text{ac}$ and let evaporation take place. We observe that coiling remains even when $D<D_\text{ac}$, see \cref{Windlass_hysteresis}, but that the drop envelope strongly deforms and that coiling rearranges to an ordered configuration (see \cref{sec:morphology}). 
The smallest measured coiling diameter is in very good quantitative agreement with the deactivation diameter $D_\text{deac}$ given in \cref{deactivation_radius_eq}.
Further evaporation, $D < D_\text{deac}$, leads to the formation of toroidal coiling held by a liquid film, which eventually snaps and leaves the coiling only bridged by fibre self-adhesion, see \cref{microcoil}.
\citet{Schulman2017} reported similar observations of ``dry coils'', prepared by winding a polymeric (SIS) fibre on the outer surface of a droplet and then removing the liquid.
Further mechanical manipulation of the evaporated sample eventually leads to irreversible uncoiling.

The subcritical nature of the in-drop buckling bifurcation thus leads to two different thresholds for coiling activation and deactivation, and to the manifestation of hysteresis: a system with $D_\text{deac} < D < D_\text{ac}$ may be either coiled or not, depending of the loading history.

%
%
%
%
%
%
%
\section{Coiling morphologies and drop deformation} \label{sec:morphology}
\begin{figure}[ht]
\begin{center}
\includegraphics[width=0.47\textwidth]{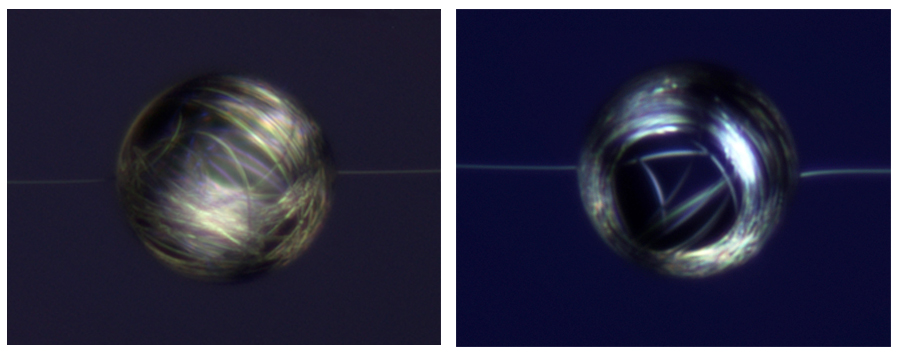}
\caption{Polarized light reveals the different coiling morphologies a TPU fibre/silicone oil drop system may adopt: disordered (phase I, left) or ordered (phase II, right). The coiling morphology depends on the precise in-drop fibre length.}
\label{Topo}
\end{center}
\end{figure}

%
%
%
%
In addition to the essentially flat shape of the plateau, \cref{fig:Undershoot} shows existence of oscillations in the moderate post-buckling regime $\Delta /(\pi \, D)=O(1)$. These oscillations have smaller amplitude than the initial undershoot peak, and decrease as the capillary force $F_\gamma$ is increased, to become essentially flat (below sensor resolution) if $F_\gamma > 300 \, EI/D^2$.  Moreover they are structured with a period $\pi D$ for the end-shortening $\Delta$, each cycle corresponding to the addition of one coil inside the drop. Numerical computations\cite{Elettro2016_theory3D} show that these cycles come from in-drop rearrangement of the coiling and that 3D and planar configurations alternate, depending on the value of the in-drop fibre length (see also supplementary video \# 1).
We indeed observe that the typical coiling morphology of a drop-on-coilable-fibre system oscillates between a fully ordered state and a fully disordered state, both extremes shown on \cref{Topo}.

\begin{figure}[th]
\begin{center}
\includegraphics[resolution=300,width=0.45\textwidth]{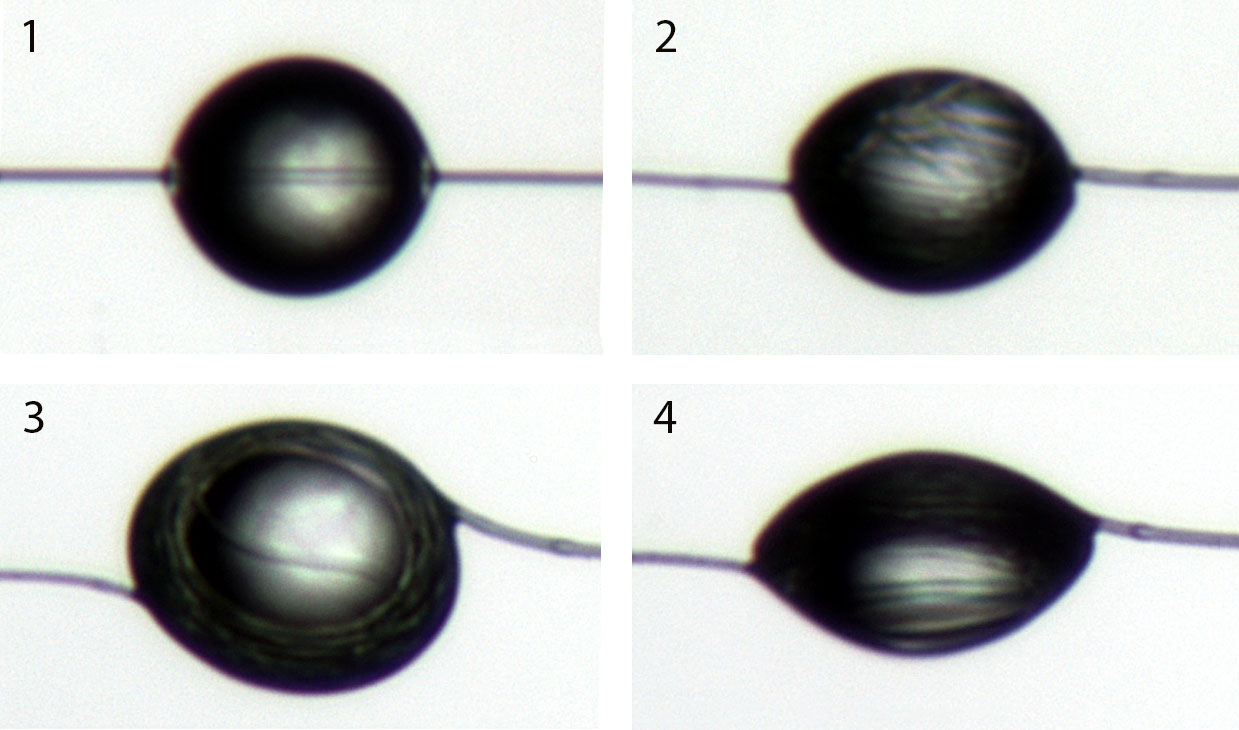}
\caption{As the in-drop fibre length increases (from 1 to 4), the initial axisymmetry of the drop may be broken, resulting in an ellipsoidal shape. (3) and (4) are rotated views of the same drop, illustrating loss of symmetry with respect to the fibre axis. We coin the axisymmetric system (1) and (2) phase I, and the ellipsoidal system (3) and (4) phase II. Here $f_\gamma=24.7 \pm 5$, $R_0=61 \pm 2 \mu$m and $r=2.5 \pm 0.4 \mu$m, and the transition from phase I to phase II occurs near 24 coils. Experimental measures of the drop deformation is reported on \cref{Topo_exp_theory}.}
\label{Axisymmetry}
\end{center}
\end{figure}

So far the liquid drop has been considered as a rigid sphere. However observations reveal that the drop envelope may undergo strong deformation when interacting with the internal coiling. Hence in the following we consider the packing of a fibre within a {\em deformable} liquid drop and study the influence of the deformability on the coiling morphology.
\paragraph*{Axisymmetry breaking}
%
%
When the fibre is straight (as in \cref{Axisymmetry}-1) and the fluid is wetting, the drop adopts a well known unduloidal shape matching the Young's contact angle in the vicinity of the fibre\citep{Carroll1976} and is symmetric with respect to the (horizontal) fibre axis for intermediate contact angles.
In presence of coiling, the fibre pushes outwards on the liquid interface with a force $F$ of magnitude $F \propto \tfrac{\partial}{\partial R} \left(\tfrac{1}{2}\tfrac{EI}{R^2}\right) = \tfrac{EI}{R^3}$ per unit length of fibre. 
This pressure field is not isotropic and may lead to asymmetric deformation of the liquid interface.
Indeed \cref{Axisymmetry} reveals the change of symmetry axis of the system as more fibre is added in the drop: the initial horizontal axis of symmetry (1) is kept as long as the coiling remains disordered (2). Eventually internal pressure from the fibre is too strong and leads to a deformed drop (3) with an axis of symmetry perpendicular to the horizontal: a localized stretch of the drop envelope represents an opportunity for the in-drop fibre to lower curvature energy. Coiling is then locked in an ordered state as more fibre is added (4).

%



\begin{figure}[ht]
\begin{center}
\includegraphics[resolution=300,width=0.45\textwidth]
{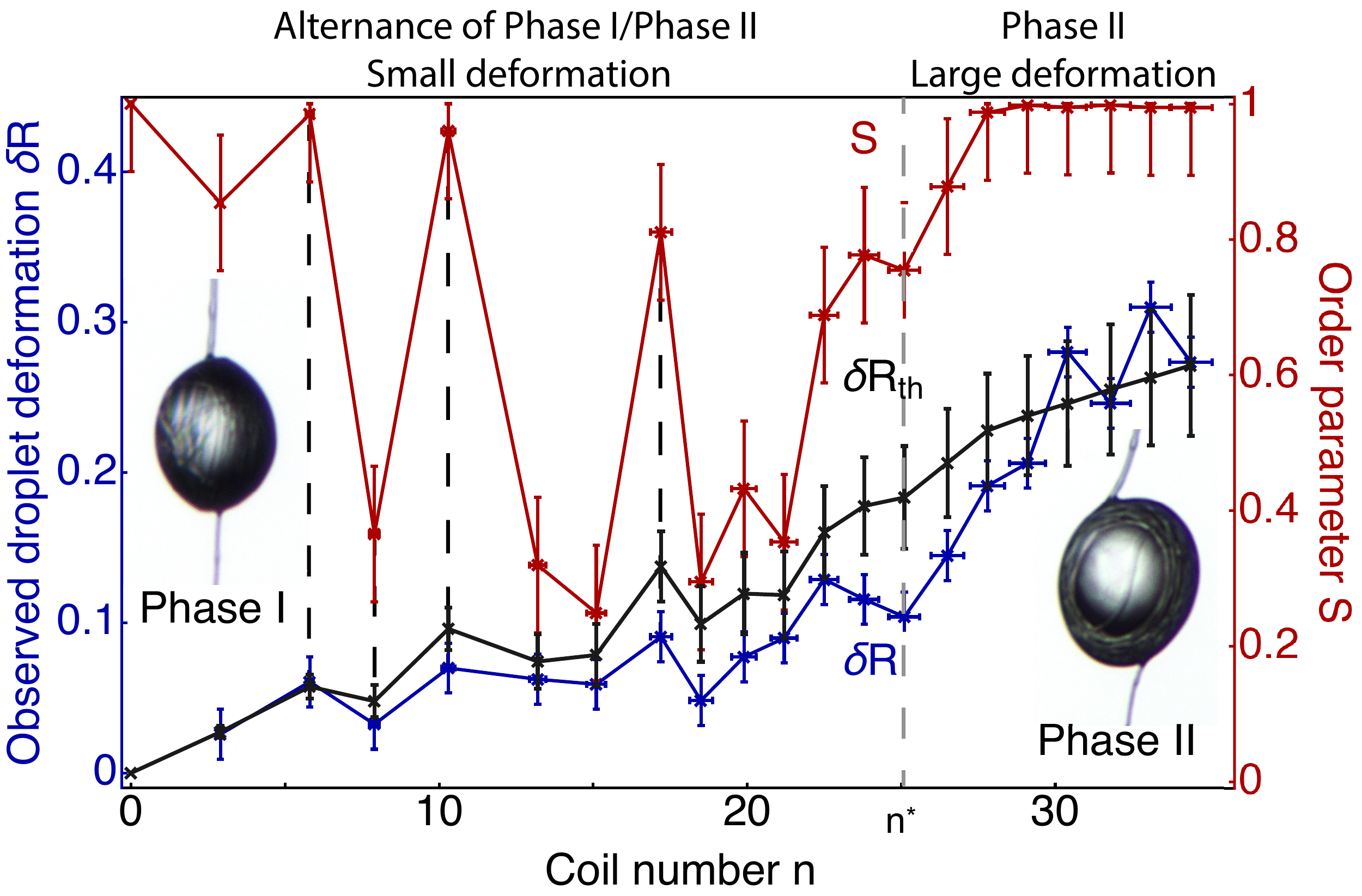}
\caption{Comparison between the measured deformation (bottom blue curve) and the theoretical prediction (middle black curve) from \cref{universal_deformation}. The top red curve is the measured order parameter $S$. Here $f_\gamma=24.7$, $R_0=61 \pm 2 \mu$m and $r=2.5 \pm 0.4 \mu$m. 
%
%
For $n<22$, coiling alternates between phase I and phase II in a regular manner, well captured by our recent numerical simulations\cite{Elettro2016_theory3D}. Drop deformation and coiling rearrangements are strongly correlated, as shown by the corresponding peaks linked by vertical dashed lines. A lock-in transition occurs at $n^\star=24$, leaving continuous ordered coiling and an highly deformed drop.}
\label{Topo_exp_theory}
\end{center}
\end{figure}

%
%
%
%
\paragraph*{Quantification of the drop deformation}
%
%
We consider a drop of initial length $2R_0$ and start increasing the in-drop fibre length $\ell$, using the apparent coil number $n=\tfrac{\ell}{2\pi R_0}$ as bifurcation parameter. 
We quantify the ordering of the coiling with an orientational order parameter $S$ commonly used in the field of liquid crystals\cite{De-Gennes1993}
\begin{equation}\label{order_parameter}
S = \langle P_2(\cos \psi) \rangle = \left \langle \frac{3 \cos^2 \left[ \psi - \langle\psi\rangle \right]-1}{2} \right \rangle
\end{equation}
with $S=0$ ($S=1$) corresponding to the fully disordered (ordered) case, and where $\psi$ is the measured angle between the fibre axis at each coil and the initial fibre axis, and $\langle \cdot \rangle$ denotes the spatial average.
In \cref{Topo_exp_theory} we plot $S(n)$ and $\delta R(n)$ where $\delta R(n)=(R(n)-R_0)/R_0$ and $2R(n)$ is the measured maximum drop length, together with the theoretically computed $\delta R_\text{th}(n)$.

\begin{figure}[ht]
\begin{center}
\includegraphics[resolution=300,width=0.45\textwidth]{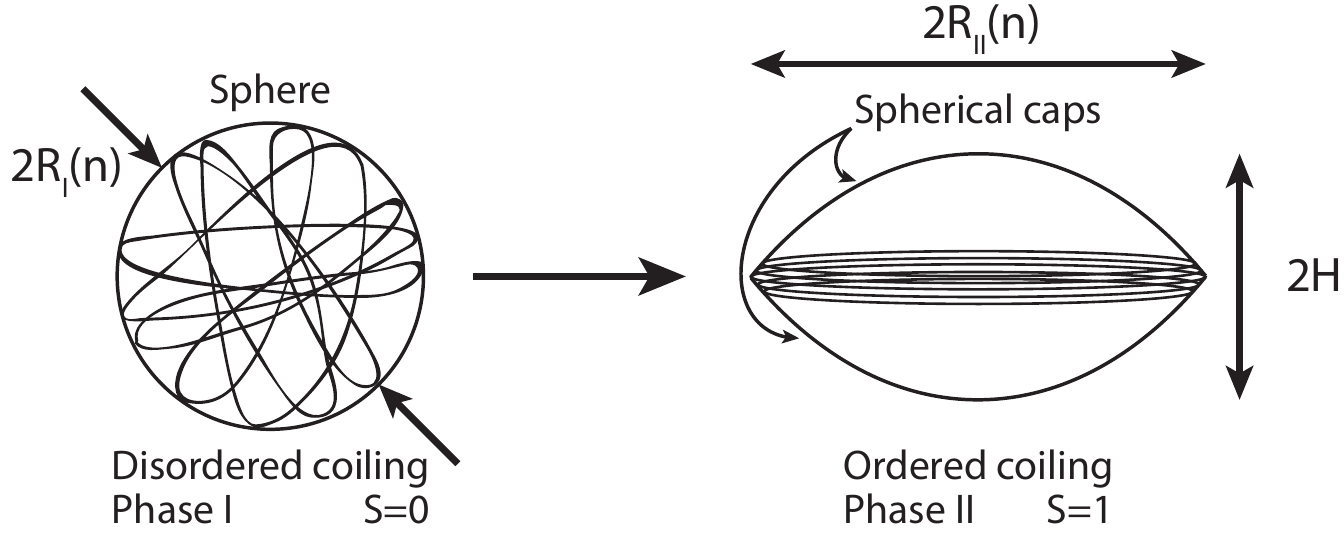}
\caption{Spherical cap model of the drop deformed by ordered in-drop fibre coiling. The extreme states are fully disordered coiling in a sphere (phase I, left) and fully ordered coiling in spherical caps (phase II, right).}
\label{Spherical_cap}
\end{center}
\end{figure}

As $n$ is increased from $n=0$, two different regimes can be identified in the system. For low values of the coil number $n$ (here typically $n \lesssim 20$), coiling is mainly disordered ($S=0$) except during short ordered intervals ($S=1$), as underlined by the strong fluctuations of order parameter during phase I (red curve in \cref{Topo_exp_theory}). These bursts of ordered coiling have been described in our recent theoretical work\cite{Elettro2016_theory3D}. 
While in phase I, the system keeps its initial horizontal axis of symmetry and we model the liquid interface as a sphere, see \cref{Spherical_cap}-left, whose radius $R(n)$ increases due to the addition of fibre volume. For small $\delta R$, conservation of total volume leads to

\begin{equation}\label{spherical_deformation}
\delta R_\text{I}(n)=\frac{R(n)-R_0}{R_0} = \frac{1}{3}\frac{V(n)-V_0}{V_0} = \frac{1}{3}\frac{\pi r^2\ell}{\frac{4}{3}\pi R_0^3} = \frac{\pi}{2}n\left(\frac{r}{R_0}\right)^2
\end{equation}
with $r$ is the fibre radius, $\ell$ the total in-drop fibre length, $V(n)=(4/3) \, \pi \, R^3(n)$ the volume enclosed by the liquid interface, and $n = \tfrac{\ell}{2\pi R_0}$ the apparent coil number.

As the coil number exceeds a threshold (this threshold is $n^\star=24 \pm 3$ for the system of \cref{Topo_exp_theory}), the system switches to continuous fully ordered coiling, phase II, see \cref{Axisymmetry}-3 and 4. Here $n^\star$ is defined experimentally as the center of the transition to constant phase II.
We model the liquid interface as two spherical caps of major (minor) axis $2R$ ($2H$), see \cref{Spherical_cap}-right.
The total energy of the system is the addition of the bending energy of the fibre $1/2 \, (EI/R^2) \, \ell$ and the surface energy of the liquid interface $2\pi \gamma (R^2+H^2)$. Minimizing this energy with regard to $R$ and $H$ under the constraint of fixed volume $\pi H \, (R^2+H^2/3)$, we find that for small surface perturbation the deformation of the drop is given by
\begin{equation}\label{ellipsoid}
\delta R_\text{II}(n)=\frac{R-R_0}{R_0} \simeq \frac{n}{4\frac{\gamma R_0^3}{E I}+3n}
\end{equation} 
We assume that for intermediate coiling morphologies, the drop deformation is the addition of the phase I deformation plus a ratio of the phase II deformation, this ratio being given by the order parameter $S$. We thus write $\delta R_\text{th}=\delta R_\text{I} + S \cdot \delta R_\text{II}$.
Combining \cref{spherical_deformation,ellipsoid} yields the theoretical drop deformation for any coiling morphology
\begin{equation}\label{universal_deformation}
\delta R_\text{th}(n) = \frac{\pi}{2}n\left(\frac{r}{R_0}\right)^2 + S(n)\times\frac{n}{4\frac{\gamma R_0^3}{E I}+3n}
\end{equation}					
\Cref{universal_deformation} is a generalization for any coil number and any order parameter of the results of Roman \& al. \cite{Roman2010}.

We tested experimentally \cref{universal_deformation} for a large number of drop-on-fibre systems. Drop deformation is measured optically as a change of length along the main fibre axis. As the fibre that lies outside the drop is taut throughout our experiments, in-drop fibre length $\ell$ is measured with the end shortening, $\ell = \Delta$.
The order parameter is measured by following optically the direction of each coil when crossing a reference line taken to be the initial fibre axis. Careful repeating of each measurement leads to errors of $\pm 0.1$, although smaller for well defined coiling ($S=0$ and $S=1$) as well as for large coiling numbers. For statistical reasons, we restrict to systems with at least $n=5$ coils. \Cref{Topo_exp_theory} shows comparison between theory and experiment for a typical measurement.The error bars on the theoretical prediction correspond to the measurement error on the order parameter S, which is an input parameter.
The present model has been tested against several drop-on-coilable-fibre systems where the radius $r$ of the TPU fibre and the initial radius $R_0$ of the silicone oil drop have been varied. The average difference between theoretical prediction and observed deformation is 15\%. Although convenient, the use of a simple setup to assess the order parameter $S$ is quite challenging. This could be improved by the use of X-ray computed tomography, which allows full 3D reconstruction in fine resolution.

We finally remark that the lock-in threshold $n^\star$ on the coil number, for which the system enters continuous phase II, could be computed by comparing the energies of both phases. Such a comparison would need an estimation of the twist energy of the system in both phases. Up to now this twist energy has been neglected  but we anticipate to be somewhat different in phases I and II, leading to an energy barrier between the two states. Using a rough estimation of this twist energy barrier as a fraction $\eta$ of the bending energy (see the Appendix 'Derivation of drop deformation and morphology transition'), we find that the threshold from phase I to phase II should happen at
\begin{equation}
n^\star = k(\eta) \, \frac{\gamma \, R_0^3}{EI} \propto \frac{R_0}{r} \, \frac{F_\gamma D^2}{EI}
\end{equation}
where $k(\eta) \sim O(1)$ is a dimensionless number and $D = 2 R_0$.
Consequently for systems moderately above the in-drop buckling threshold $F_\gamma =\pi^2 \, EI/D^2$, the transition to ordered coiling will happen for a coil number of order 10 to 100. In contrast, systems far above the buckling threshold may have their transition to ordered coiling prevented as the system may reach the close-packing limit. Estimating this limit\cite{Katzav2006} as $n \propto R^2/r^2$, we see that a system where the fibre radius $r$ satisfies $r<r_\text{c} \propto \sqrt{\frac{\gamma}{E}R_0}$ 
 will likely not experience the transition to ordered coiling.



%
%
%
%
%
%
\section{Conclusion}
%
%
%
%
In conclusion, we showed that the possibility of transferring wetting energy into mechanical energy leads to events of negative stiffness regimes in drop-on-coilable-fibre systems. 
Consequences include force undershoots in displacement-controlled setups, and mechanical hysteresis in force-controlled setups.
In both cases, quantitative agreement between experiments, theory and numerics has been reached.
We observed that further occurrences of negative stiffness regimes are linked to changes in the coiling morphology, and showed that a lock-in transition may eventually occur, underlining the link between the in-drop fibre length and the drop deformation.
Drop-on-coilable-fibre systems thus open new possibilities as complex actuators in light of their unconventional mechanical response. For instance, accurate control may yield new routes to 3D microfabrication and reconfigurable coil-to-cage devices in liquid environments.

%
%
%
%
%
%
%
\section*{Acknowledgements}
%
%
%
%
We gratefully thanks Natacha Krins, David Grosso, C\'edric Boissi\`ere, Sinan Haliyo, and Camille Dianoux for helpful discussions.
The present work was supported by the french Agence Nationale de la Recherche, grants ANR-09-JCJC-0022-01 and ANR-14-CE07-0023-01, `La Ville de Paris, Programme Emergence', The U.K. Royal Society, grant IE130506 of the International Exchanges Scheme, the CNRS through a PEPS PTI program, the US Air Force, AFOSR grant FA9550-12-1-0294, the European Research Council, grant SP2-GA-2008-233409, and the Conicyt/Fondecyt post-doctoral n$^{\circ}$ fellowship 3160152.
%
%
%
%
%
%

\section*{Appendix: Materials and Methods}
%

We prepare microfibres with Thermoplastic PolyUrethane (TPU) from BASF (reference 1185A). We use melt spinning to draw fibres: after TPU pellets are molten on a hot plate at 230\si{\degree}C, a small amount of liquid TPU is then quickly stretched, followed by quenching at room temperature. The obtained fibres have a radius that lies between 1 and 10 \si{\micro\meter}, depending on the stretching speed. The goal radius is usually achieved within a 2 \si{\micro\meter} range. The local radius variations are at most by $10\%$ over an extended region of several thousands radii, which corresponds to the typical maximum in-drop fibre length.
The fibre and drop diameters are measured optically with a 3 megapixels Leica DFC-295 camera mounted on a Leica macroscope (VZ85RC, 400x zoom, 334 nm/pixel picture resolution) and a remote-controlled micro-step motor.
We either used a Phlox 50x50 mm backlight (60000 Lux) or an optical fibre with LED lamp (Moritex MHF-M1002) with circular polarizer. The use of polarized light strongly enhances the visibility of the fibre inside of the drop due to birefringence of TPU microfibres.
We measured contact angles by superimposing optical images of drops on fibres to corresponding calculated profiles, and found $\theta_Y=23\pm2$\si{\degree} for the TPU/silicone oil setup and $\theta_Y=19\pm2$\si{\degree} for the TPU/ethanol setup. We used $\gamma=21.1 \,$\si[per-mode=symbol]{\milli\newton\per\meter} for silicone oil/air interface and $\gamma=22.1 \,$\si[per-mode=symbol]{\milli\newton\per\meter} for ethanol/air interface. For evaporation-controlled experiments, we use a megasonic transducer activated at 1.6 \si{\mega\hertz} (Beijing Ultrasonics) to produce a cloud of micronic droplets, with sizes in the 3--5 \si{\micro\metre} range (inferred optically) and controllable outflow.
Due to the low intensity of the forces (typically hundreds of nN for our microfibres), we use capacity deflection force sensors (FemtoTools FT-FS100, 5 nN-100$\mu$N range at 20 Hz). While highly sensitive, these sensors have the drawback of drifting slowly with time around 10 \si[per-mode=symbol]{\nano\newton\per\minute}. This only adds slightly to the measurements errors in rapidly performed tests, typically less than five minutes. In longer tests, the drift can lead to substantial offsets and affects our ability to evaluate absolute force values. Consequently the theoretical value of the plateau tension, given by \cref{equa:T-analytic}, is used to tare the experimental data on \cref{fig:Undershoot}. However, the drift is compensated for graphical purpose only, and does not affect measurements in \cref{Undershoot_vs_Lw}-bottom, as force undershoots are relative quantities.
The force sensor is mounted on a linear micro-positioner (SmarAct SLC-1730, repeatability 0.5 $\mu$m) and all the tests are displacement-controlled and performed at a quasi-static speed of 12 $\mu$m/s.
Young's modulus of the fibres are measured through tensile tests and found to be $17\pm2$ MPa and do not depend on conditions of preparation. By brushing a fibre with a viscous silicon oil drop (Rhodorsil 47V1000) hanging from a syringe tip, we obtain an array of drops of different sizes. The activation of in-drop buckling is tested by compression of the resulting sample.

%
%
%
%
\section*{Appendix: Derivation of drop deformation and morphology transition}
%
%
%
%
%
%

We consider the ordered phase II and detail the calculations for the deformation of the drop envelope under the spherical cap model, see \cref{Spherical_cap}.
The total potential energy of the system is the addition of the bending energy of the fibre $E_b(R)=1/2 \, (EI/R^2) \, \ell$, the twist energy of the fibre $E_t$, and the surface energy of the two spherical caps $E_S(R,H)=2\pi \gamma (R^2+H^2)$.
We have a constraint of fixed liquid volume expressed as $(1/3) \,\pi H \, (3R^2+H^2)=(4/3) \, \pi R_0^3$, where $R_0$ is the radius of the undeformed spherical drop.
As the twist energy $E_t$ is difficult to evaluate we first do not take it into account, and introduce the lagrangian of the system as $\mathcal{L}(R,H)=E_b(R)+E_S(R,H) - \lambda(1/3 \,\pi H[3R^2+H^2]- 4/3 \, \pi R_0^3)$, where $\lambda$ is the Lagrange multiplier associated to volume conservation and is identified to the pressure inside the drop.
The equilibrium of the system is then given as a stationary point of $\mathcal{L}$:
\begin{subequations}
\label{sys12}
\begin{align}
\label{Lagrangian_H}
\frac{\partial \mathcal{L}}{\partial H} & = 0 = 4\pi\gamma H - \lambda \pi (R^2+H^2)\\
\label{Lagrangian_R}
\frac{\partial \mathcal{L}}{\partial R} & = 0 = -\frac{EI\, \ell}{R^3}  + 4\pi\gamma R - 2 \pi R \lambda H
\end{align}
\end{subequations}
We work under the assumption that the deformation of the interface is small, that is when $R$ and $H$ are both near $R_0$ and the pressure $\lambda$ is near $2\gamma/R_0$.
We then write $R=R_0(1+\epsilon)$ and $H=R_0 + \epsilon H_1$ with $|\epsilon| \ll 1$. Volume conservation yields $H_1=-R_0$, that is $H=R_0(1-\epsilon)$. System (\ref{sys12}) is then solved and we obtain $\lambda=(2\gamma/R_0) (1-\epsilon)$ and
\begin{equation}
\epsilon  = \frac{\ell \, EI }{8 \pi \gamma R_0^4 + 3 \ell \, EI} = \frac{n \, EI }{4 \gamma R_0^3 + 3 n \, EI}
\label{eq13}
\end{equation}
where we see that for $\epsilon$ to be small, we need the capillary force $\gamma R_0$ to be large compared to the bending force $EI/R_0^2$, and the coil number $n= \ell /(2\pi R_0)$ not to be too large. 

We conclude that when twist energy $E_t$ is neglected, we have a unique equilibrium at $\epsilon > 0$ (ordered coiling) and no equilibrium at $\epsilon=0$ (disordered coiling), but we reckon that taking the twist energy into account might change the situation and stabilize the $\epsilon=0$ state.
The twist energy is larger in the ordered coiling configuration than in the disordered coiling configuration for the following reason. In the experiments the fibre is held at both extremities, therefore imposing a constraint of zero Link, that is zero end rotation of the ends\cite{fuller:1978,Heijden2007}. 
As a rod deforms in space the Writhe is a real number which measures the circumvolution of the center line in 3D. The more windings, the higher the Writhe. An estimation of the Writhe is obtained by (i) looking at the rod from a given point of view and projecting the rod shape on a plane perpendicular to the view axis, (ii) counting the number of crossings on the projection, and (iii) start again with every view axis and average the result.
The Twist is the integral along the rod of the local twist.
An important feature of twisted rod mechanics is that, if both extremities of the rod are held fixed, as the rod deforms in space its Writhe and its Twist change but at all times the sum of the Twist and the Writhe stays constant, equal to the Link \cite{Dennis2005}. 
In the disordered coiling configuration we estimate the Writhe to be small due to a statistical balance of positive and negative crossings. Hence the Twist, and therefore the twist energy, is to be small. However in the ordered coiling configuration the Writhe is almost that of a regular spool with $n$ turns, Writhe $\approx n$. We then have Twist $\approx -n$, in order for the sum to be zero. The twist energy of the ordered coiling configuration is then larger than the twist energy of the disordered coiling configuration. 
It is difficult to be more quantitative without performing complete numerical simulations. Here we simply estimate the difference in twist energy is equal to a fraction $\eta$ of the bending energy $E^{II}_t-E^I_t \simeq (1/2) \eta \, (EI/R_0^2) \ell$, with $\eta \sim O(0.1)$. This estimation follows from numerical results from \citet{Stoop2011} and \cite{Elettro2016_theory3D}.

We now consider the total potential energy $V=E_b+E_t+E_s$ in both disordered coiling, $V^I$, and ordered coiling, $V^{II}$.
We evaluate $V^I$ at $R=H=R_0$ and $V^{II}$ at $R=R_0(1+\epsilon)$ and $H=R_0(1-\epsilon)$ when $\epsilon$ given by \cref{eq13}. We then compute the first order of the difference $V^{II}-V^I$:
\begin{equation}
V^{II}-V^I \simeq \frac{2 \pi \, EI \, n}{R_0} \left( \frac{\eta}{2} - \frac{n \, EI }{4 \gamma R_0^3 + 3 n \, EI} \right)
\end{equation}
For small $n$, twist energy makes the disordered configuration favorable ($V^{II}-V^I>0$), but as $n$ reaches a threshold $n^\star = \frac{4\eta}{2-3\eta} \, \frac{\gamma R_0^3}{EI}$ the ordered configuration becomes favorable ($V^{II}-V^I<0$).






\bibliography{Undershoot_and_morphology} 
\bibliographystyle{rsc} 

\end{document}